\documentclass[conference,a4paper]{IEEEtran}
\IEEEoverridecommandlockouts

\usepackage{cite}
\usepackage{amsmath,amssymb,amsfonts,threeparttable}
\usepackage{algorithmic}
\usepackage{graphicx}
\usepackage{textcomp}
\usepackage{hyperref}
\usepackage{xcolor}
\usepackage{flushend}
\usepackage{fancyhdr}

\def\BibTeX{{\rm B\kern-.05em{\sc i\kern-.025em b}\kern-.08em
		T\kern-.1667em\lower.7ex\hbox{E}\kern-.125emX}}
	
\fancypagestyle{firstpage}{
	\setlength{\headheight}{3.5\normalbaselineskip}
	\setlength{\headsep}{1.5\normalbaselineskip}

	\fancyhead[C]{%
		\footnotesize%
		To be presented at the \textit{2026 IEEE International Conference on Visual Communications and Image Processing}, Singapore, December 2026 \\
		\vspace{0.8\normalbaselineskip}%
		\scriptsize
		\copyright{} 2026 IEEE. Personal use of this material is permitted. Permission from IEEE must be obtained for all other uses, in any current or future media, including reprinting/republishing this material for advertising or promotional purposes, creating new collective works, for resale or redistribution to servers or lists, or reuse of any copyrighted component of this work in other works.%
	}
	\fancyfoot{}
}
	
\begin{document}
	
	\title{Inter-Frame Predictive Contour Coding for Semantic Map Compression}
	
	\author{\IEEEauthorblockN{Runyu Yang}
		\IEEEauthorblockA{\textit{Simon Fraser University}\\
			Burnaby, BC, Canada \\
			runyuy@sfu.ca}
		\and
		\IEEEauthorblockN{Hyomin Choi}
		\IEEEauthorblockA{\textit{InterDigital AI Lab} \\
			Los Altos, CA, USA \\
			hyomin.choi@interdigital.com}
		\and
		\IEEEauthorblockN{Fabien Racap\'e}
		\IEEEauthorblockA{\textit{InterDigital AI Lab} \\
			Los Altos, CA, USA \\
			fabien.racape@interdigital.com}
		\and
		\IEEEauthorblockN{Ivan V. Baji\'c}
		\IEEEauthorblockA{\textit{Simon Fraser University}\\
			Burnaby, BC, Canada \\
			ibajic@ensc.sfu.ca}
	}
	
	\maketitle
	\thispagestyle{firstpage}
	
	\begin{abstract}
		Dynamic semantic maps have become an important intermediate representation for many computer vision applications. Efficient lossless compression of such maps is essential for reducing storage and transmission costs. Existing contour-based semantic map compression methods mainly exploit spatial redundancy within individual frames while largely ignoring the strong temporal correlation between adjacent semantic maps. In this paper, we propose an inter-frame predictive contour coding framework for lossless semantic map compression. The proposed framework first predicts blob positions and label indexes from the reference frame and then performs contour prediction using corresponding chain-code units in the reference contours. A mode selection mechanism is further employed to choose between intra and inter coding according to the estimated coding cost. By exploiting temporal contour redundancy while preserving the advantages of contour-based representation, the proposed framework significantly improves coding efficiency. Experimental results on the VSPW dynamic semantic segmentation dataset demonstrate that the proposed method consistently outperforms state-of-the-art lossless semantic map compression methods. Compared with the recent context-adaptive extended chain coding framework, the proposed inter-frame coding achieves an average bitrate reduction of 3.0\% without significant increase in the average encoding time, demonstrating an effective trade-off between compression performance and computational complexity. The source code is made publicly available at \url{https://github.com/InterDigitalInc/LosslessSegmentationMapCompression}.
	\end{abstract}
	
	
	\section{Introduction}
	\label{sec:intro}
	Semantic maps assign a semantic label to every pixel and have become an essential intermediate representation for numerous computer vision applications, including autonomous driving, video scene understanding, visual tracking, and embodied intelligence. With the rapid development of video semantic segmentation, large-scale dynamic semantic maps are increasingly generated, transmitted, and stored. Efficient lossless compression of dynamic semantic maps is therefore important for reducing transmission bandwidth and storage requirements while preserving semantic information.
	
	Unlike natural images, semantic maps contain a limited number of labels and sharp object boundaries. These characteristics make conventional image and video compression methods, such as FLIF \cite{sneyers2016flif} and HEVC Screen Content Coding (HEVC-SCC) \cite{xu2015overview}, unable to fully exploit their structural properties. Recently, contour-based coding has demonstrated superior performance by representing semantic maps using object boundaries instead of pixel values. In particular, a context-adaptive chain coding based framework \cite{yang2026context} achieves state-of-the-art compression efficiency for static semantic maps. It first divides the semantic map to semantic regions with the same label, referred to as \emph{blobs}, then exploits contour structure and shared contours between adjacent blobs. However, it just focuses on intra-frame coding and does not exploit the strong temporal correlation between adjacent semantic maps in video sequences. Zhu \emph{et al.} \cite{zhu2015highly} proposed a highly efficient method to exploit temporal correlations among contours using chain-based motion estimation and motion compensation. But they operated on binary video and did not consider shared contours between adjacent blobs.
	
	To address this problem, this paper proposes an inter-frame predictive contour coding framework for lossless semantic map compression. Instead of performing motion estimation on pixels or blocks, the proposed method directly exploits the temporal redundancy of blob contours. Blob positions and label indexes are first predicted from the reference frame. Then, contour prediction is performed by searching corresponding chain-code units in the reference contour, retaining efficient skip coding for shared contours between adjacent blobs. Finally, the mode selection is performed between intra and inter coding according to rate estimation. This design effectively captures contour similarity across adjacent frames while preserving the advantages of contour-based representation.
	
	Experimental results on the VSPW dynamic semantic segmentation dataset demonstrate that the proposed framework consistently outperforms state-of-the-art lossless compression methods, including FLIF, HEVC-SCC, and recent contour-based semantic map compression methods. Compared with the intra-only framework, the proposed inter-frame coding further reduces the average bitrate while maintaining low computational complexity, demonstrating the effectiveness of exploiting temporal contour redundancy in dynamic semantic maps.
	
	\begin{figure*}[th!]
		\centering
		\begin{minipage}[b]{1\linewidth}
			\centering
			\centerline{\includegraphics[width=\linewidth]{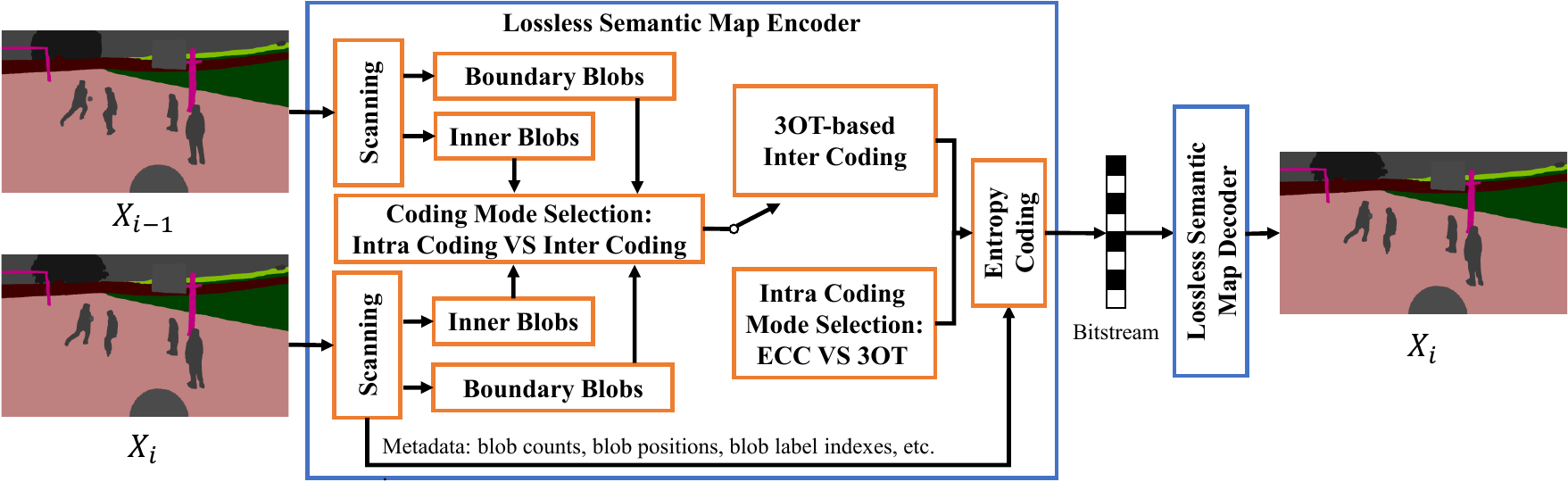}}
		\end{minipage}
		\caption{System schematic of the overall lossless semantic map compression framework.}	
		\label{fig:overview}
	\end{figure*}
	
	The main contributions of this paper are summarized as follows:
	\begin{itemize}
		\item We propose an inter-frame predictive contour coding framework for dynamic semantic map compression.
		\item A contour prediction scheme based on position, label, and chain codes prediction is developed to exploit temporal redundancy.
		\item Experimental results demonstrate consistent bitrate reduction over state-of-the-art lossless coding methods.
	\end{itemize}
	
	\section{Proposed Framework}
	\label{sec:Proposed}
	
	An overview of the proposed lossless semantic map compression framework is illustrated in Fig.~\ref{fig:overview}. The semantic map is first divided into 2D slices and then compressed slice by slice. The intra coding follows a chain coding-based semantic map compression method \cite{yang2026context}. For encoding the label of each blob, a single global label palette is constructed from all frames according to the raster-scan order and reused for all frames in the sequence, ensuring that a given label index refers to the same semantic class. The label will be restored at the decoder by using the same shared label palette. The inter coding uses the previous coded map as the reference frame and mainly consists of two key steps: (1) inter coding of blob position and label index; and (2) inter coding of chain codes. It first scans the semantic map to get all blobs, following the counterclockwise order from the top-left pixel to get boundary blobs and the raster-scan order for the inner blobs. The position of each blob is the first scanned pixel in the blob. It then finds predictable blobs to predict blob positions and label indexes. For each predictable blob, both intra coding and inter coding methods are evaluated, and the encoder selects the method that yields the lower total number of bits to represent the contour. The selected coding mode is signaled in the bitstream using a mode flag. To achieve better coding performance, we employ an adaptive arithmetic coder to encode all the syntax elements and estimate rates.
	
	\subsection{Inter coding of blob position and label index}
	
	Because blobs in adjacent maps are similar, most blobs in the previously coded map can find the corresponding blob with same label index in current map, the bitrate of the label index can be saved by setting a flag to show whether the corresponding blob in current map has been found. The corresponding blob is the blob with the same label index and has the closest position to the blob. If the corresponding blob is found, it needs to transmit the difference between two blob position coordinates using order-0 Exponential-Golomb coding and does not need to transmit the label index. The other positions and label indexes are transmitted by the intra coding method \cite{yang2026context}.
	
	\begin{figure}[th!]
		\centering
		\begin{minipage}[b]{1\linewidth}
			\centering
			\centerline{\includegraphics[width=\linewidth]{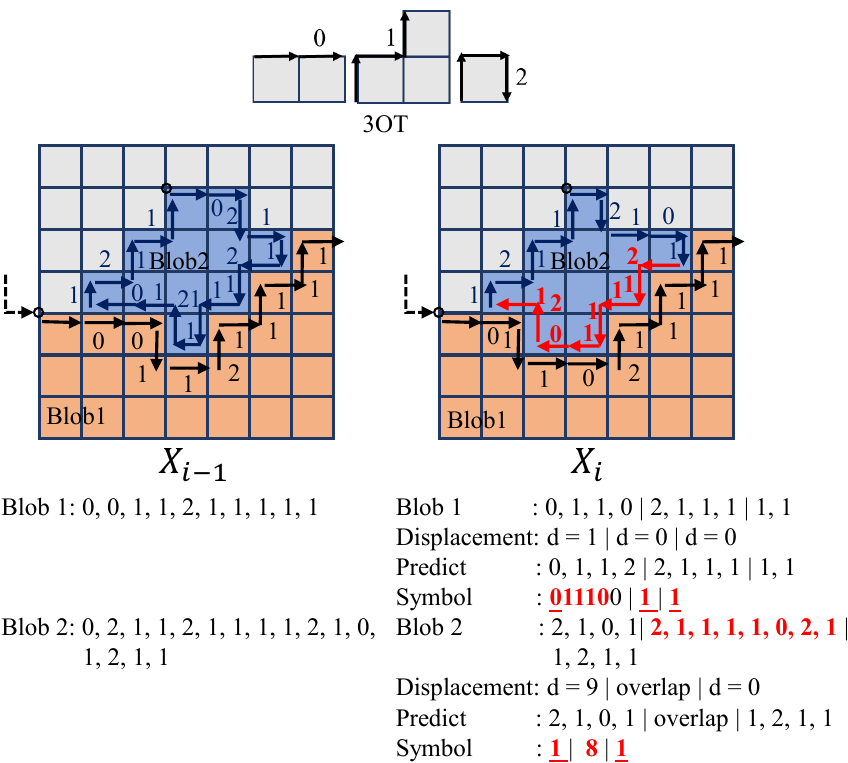}}
		\end{minipage}
		\caption{Example of the proposed inter coding method of chain coding.}	
		\label{fig:inter}
	\end{figure}
	
	\subsection{Inter coding of chain codes}
	
	We use the chain code to represent the contour of each blob. The starting point of the chain code is the same as the position of the blob. For each scanned blob in current map, if the blob’s position is predicted by a blob in the reference frame, the chain codes of the blob in the reference frame are used as reference chain codes. Fig.~\ref{fig:inter} shows an example of the proposed inter coding method using the three-orthogonal chain code (3OT) \cite{cruz2005compressing}. The 3OT chain code consists of three symbols: symbol 0 indicates that the contour direction is preserved, symbol 1 indicates a change in direction that differs from the previous change, and symbol 2 is used otherwise.
	
	The chain codes without the already encoded contour are first divided into a number of prediction units based on a given unit length (the unit length is set to 4 in Fig.~\ref{fig:inter}). For each prediction unit in current blob, disparity estimation is performed based on the same position in reference chain codes. The predict flag is used for each chain code to show whether the prediction is correct. If the prediction is wrong, chain coding is performed following the corresponding predict flag. To efficiently code the predict flag, a skip flag is set to show whether the whole prediction unit is predicted correctly. The disparity displacement $d$ is selected by exhaustive search over all candidate displacements within the search range setting. For each candidate $d$, the estimated coding cost is the sum of: (1) the order-0 Exponential-Golomb coding for $d$, (2) the skip flag, (3) the predict flag for each non-skipped chain code symbol, and (4) mispredicted chain code symbols. The displacement with minimum total estimated rate is selected.
	
	If the chain code reaches the already encoded contour, the skip coding is used.	The skip coding is also used in intra coding, which has two distinct cases: a complete skip mode and a partial skip mode. The complete skip mode is signalled using a flag, e.g., called \emph{complete skip flag}. When the flag is set to 1, the whole contour from the current position towards the starting position is skipped. When the flag is set to 0, the mode is inferred to be partial skip. In this case, a run-length value is coded, which corresponds to the lengths in symbols of the skipped contour (e.g., ``Symbol: 8'' in Fig.~\ref{fig:inter}). The inter coding of the contour resumes at the next location indicated by this length and the shared contour.

	\subsection{Decoding process}
	
	\begin{figure}[!t]
		\centering
		\begin{minipage}[b]{0.78\linewidth}
			\centering
			\centerline{\includegraphics[width=\linewidth]{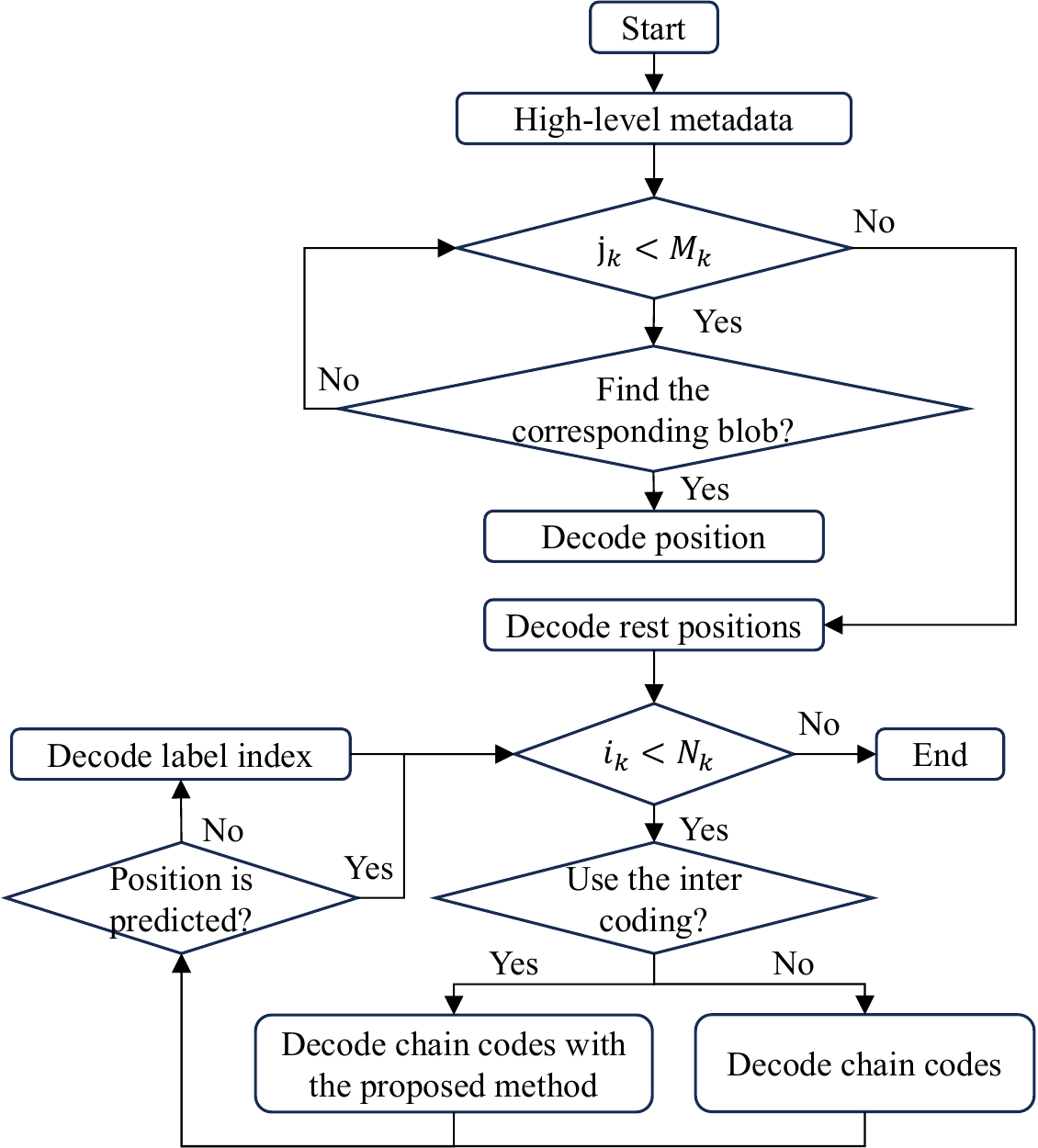}}
		\end{minipage}
		\caption{Overview of decoding process of the proposed framework.}	
		\label{fig:decoding}
	\end{figure}
	
	Since the bitstream should be self-decodable, not only the chain codes, as well as the high-level parameters such as the number of blobs are coded in the bitstream. Fig.~\ref{fig:decoding} depicts overall decoding flow. High-level parameters are parsed first. $N_k$ is the total number of coded blobs. $M_k$ is the total number of blobs in the reference frame. $i_k$ counts the decoded blob numbers over the decoding process and $j_k$ counts the scanned blobs in the reference frame. Decoding process continues until $i_k$ reach to $N_k$.  
	
	To decode all blobs, positions are parsed first along with chain coding followed by label index coding. If the inter coding flag is true for use of the proposed inter coding method. Otherwise, using the intra coding method to obtain the contour and reconstruct a blob.
	
	\begin{table}
		\caption{Information of the tested VSPW sequences.}
		\begin{center}
			\begin{tabular}{c|c|c|c}
				\hline
				Test Sequence & Resolution & Frames & Camera Motion \\
				\hline
				112  & $1280 \times 720$ & 136 & Static \\
				\hline
				127 & $1280 \times 720$ & 121 & Dollying Slowly \\
				\hline
				231 & $1280 \times 720$ & 136 & Moving Slowly \\
				\hline
				1296  & $1920 \times 1080$ & 45 & Static \\
				\hline
				1643 & $1920 \times 1080$ & 45 & Moving Fast \\
				\hline
				2097 & $1920 \times 1080$ & 45 & Moving Fast \\
				\hline
			\end{tabular}
			\label{information}
		\end{center}
	\end{table}
	
	\begin{figure}[!t]
		\centering
		\begin{minipage}[b]{1\linewidth}
			\centering
			\centerline{\includegraphics[width=\linewidth]{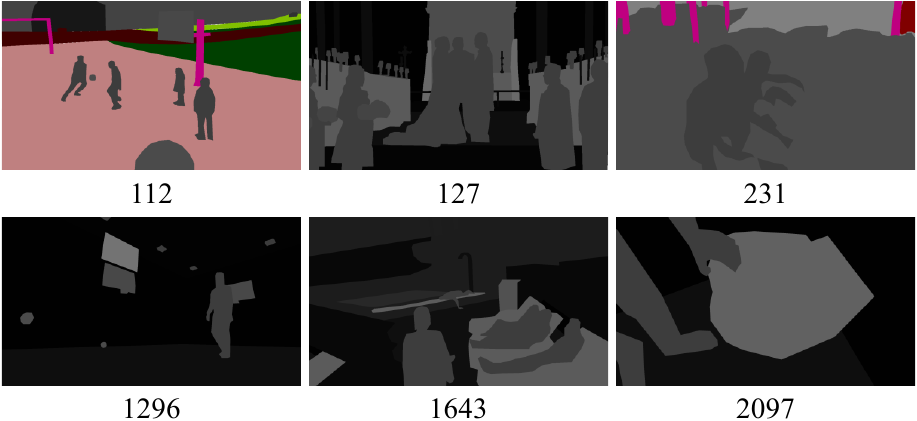}}
		\end{minipage}
		\caption{Sample maps from each test sequence.}	
		\label{fig:visual}
	\end{figure}
	
	\section{Experimental Results}
	\label{sec:experiment}
	\subsection{Experimental Settings}
	\label{Experimental Settings}
	
	\begin{table*}
		\caption{Performance of our framework compared with FLIF, SCM-7.0, and \cite{yang2026context} in the VSPW coding.}
		\begin{center}
			\begin{tabular}{c|c|c|c|c|c|c|c|c|c}
				\hline
				Test Sequence&FLIF&SCM-7.0*&SCM-7.0&CC-SMC*&CC-SMC&\cite{yang2026context}&Ours& Gain of Inter&Ours w/o skip coding\\
				\hline
				112&269805&354154&234689&173373&150269&133810&127567&-4.67\%&128825\\
				127&389373&528348&380226&272718&258553&214084&205317&-4.10\%&205345\\
				231&265486&408058&322969&182690&180275&143971&141793&-1.51\%&141803\\
				1296&83681 &124523&70984&44970&41189&33278&31817&-4.39\%&31900\\
				1643&187640&297341&211207&118621&118570&89948&89312&-0.71\%&89310\\
				2097&67628&135504&80927&39845&39772&28251&28162&-0.32\%&28162\\
				\hline
				Avg Rate (Byte)&210602&307988&216834&138703&131438&107224&103995&&104224\\
				\hline
				Avg Enc Time (s)&48.24&353.97&1097.63&3.85&73.17&2.38&2.38&&2.47\\
				\hline
				Avg Dec Time (s)&5.30&5.83&4.64&0.58&0.57&0.48&0.73&&0.77\\
				\hline
			\end{tabular}
			\begin{tablenotes}
				\item[1] The method with * means without inter coding. 
			\end{tablenotes}
			\label{results}
		\end{center}
	\end{table*}
	
	In the experiments, the unit length was set to 12, while the search range for the reference unit in inter coding was fixed at 32. The proposed framework was compared with SCM-7.0, FLIF, CC-SMC, and the intra-only version of the proposed framework \cite{yang2026context}. SCM-7.0, the reference software for HEVC-SCC \cite{xu2015overview}, represents one of the state-of-the-art screen content coding schemes. To enable the palette mode in SCM-7.0, each semantic map was converted into the YUV420 format, with the U and V channels set to zero. FLIF is one of the most efficient lossless image compression formats \cite{sneyers2016flif}. CC-SMC is a recent chain coding-based semantic map compression method containing inter coding tools \cite{yang2024cc}.
	
	Experiments were conducted on the VSPW dataset \cite{miao2021vspw}, the first large-scale dynamic semantic segmentation dataset covering multiple scenes at a frame rate of 15 fps. Since the effectiveness of inter coding depends on motion characteristics, six validation sequences exhibiting different camera motion patterns were selected for evaluation. Detailed information about the test sequences is provided in Table~\ref{information} and sample maps from each test sequence are shown in Fig.~\ref{fig:visual}.
	
	For SCM-7.0, both the All-Intra (AI) and Low-Delay (LD) configurations were evaluated. Similarly, the CC-SMC framework was tested in both intra-only and the inter coding mode.
	
	\subsection{Results}
	\label{video results}
	
	Table~\ref{results} compares the proposed framework with FLIF, SCM-7.0, and CC-SMC on the VSPW dataset. The proposed framework consistently achieves the lowest bitrate among all compared methods. Compared with its intra-only version, the proposed method further improves the coding efficiency. Specifically, the proposed framework reduces the average bitrate from 107,224 bytes to 103,995 bytes, corresponding to an average bitrate reduction of approximately 3.0\%. The improvement is consistently observed across all test sequences, indicating that the proposed inter-frame predictive contour coding effectively exploits temporal correlation between adjacent semantic maps. 
	
	As shown in the second column on the right of Table~\ref{results}, the inter coding achieves bitrate reductions ranging from 0.32\% to 4.67\%, with a per-sequence average of 2.62\%. Larger gains are obtained for sequences with static or slowly moving cameras (Sequences 112, 127, and 1296), where adjacent semantic maps exhibit stronger temporal consistency. In contrast, the gains become smaller for sequences with fast camera motion (Sequences 1643 and 2097), since rapid motion decreases the similarity between consecutive contours and therefore limits the prediction accuracy. Nevertheless, the proposed inter coding still provides consistent bitrate savings in all tested scenarios. The last column of Table~\ref{results} highlights the impact of the skip coding adopted in inter coding. This result demonstrates the importance of retaining the skip coding to improve coding efficiency.
	
	In terms of computational complexity, the proposed framework also demonstrates favorable performance. The average encoding time is similar to that of the intra-only version. Although the average decoding time slightly increases from 0.48 s to 0.73 s due to the additional inter prediction process, the decoding complexity remains significantly lower than that of SCM-7.0 while maintaining superior compression performance. These results indicate that the proposed framework achieves a favorable trade-off between coding efficiency and computational complexity.
	
	\section{Conclusion}
	This paper proposed an inter-frame predictive contour coding framework for lossless semantic map compression. By exploiting the temporal correlation between adjacent semantic maps, the proposed framework performs predictive coding of blob positions, label indexes, and contour chain codes. A mode selection strategy was further introduced to choose between intra and inter coding according to the estimated coding cost. Experimental results on the VSPW dataset demonstrated that the proposed method consistently outperforms state-of-the-art lossless semantic map compression methods. Compared with the recent context-adaptive extended chain coding framework, the proposed method achieved an average bitrate reduction of 3.0\% without significant increase in the average encoding time. These results demonstrate that exploiting temporal contour redundancy is an effective solution for improving the compression efficiency of dynamic semantic maps. Future work includes extending the proposed framework to dynamic 3D semantic map compression.
	
	\bibliographystyle{IEEEbib}
	\bibliography{refs}
	
\end{document}